\documentclass[sort&compress]{aipproc} 
\layoutstyle{6s}
\usepackage{citesort}
\usepackage{amssymb}
\usepackage{amsmath}
\usepackage{graphicx}

\title{Lessons about likelihood functions \\from nuclear physics}

\classification{29.85.Fj, 07.05.Kf}
\keywords{likelihood, Student t distribution, long-tailed likelihood functions, systematic uncertainty, inconsistent data, outliers, robust analysis, least-squares analysis}
\author{Kenneth M. Hanson}
 {
  address = {\centering{T-16, Nuclear Physics, Los Alamos National Laboratory \\Los Alamos, New Mexico, USA  087545 \\{kmh@lanl.gov}}},
  email = {kmh@lanl.gov}
 }
\date{}

\def\onehalf{{\scriptstyle \frac{\raise.1ex\hbox{$\scriptstyle 1$}}{\lower.3ex\hbox{$\scriptstyle 2$}} }}
\def\smallfrac#1#2{{ \frac{\lower.5ex\hbox{$\small #1$}}{\raise.3ex\hbox{$\small #2$}} }}
\def\Smallfrac#1#2{{ \frac{\lower.5ex\hbox{$\footnotesize #1$}}{\raise.3ex\hbox{$\footnotesize #2$}} }} 
 
\def\given{\,|\,}   

\def\mvec#1{\mbox{\boldmath $\bf#1$}}   

\begin{abstract}
Least-squares data analysis is based on the assumption that the normal (Gaussian) distribution appropriately characterizes the likelihood, that is, the conditional probability of each measurement $d$, given a measured quantity $y$, $p\,(d \given y)$. On the other hand, there is ample evidence in nuclear physics of significant disagreements among measurements, which are inconsistent with the normal distribution, given their stated uncertainties. In this study the histories of 99 measurements of the lifetimes of five elementary particles are examined to determine what can be inferred about the distribution of their values relative to their stated uncertainties.  Taken as a whole, the variations in the data are somewhat larger than their quoted uncertainties would indicate.  These data strongly support using a Student t distribution for the likelihood function instead of a normal. The most probable value for the order of the t distribution is 2.6~$\pm$~0.9. It is shown that analyses based on long-tailed t-distribution likelihoods gracefully cope with outlying data.
\end{abstract}

\begin{document}
\maketitle

\section{Introduction}

The likelihood plays a central role in any inference process. The likelihood is the conditional probability $p\,(d \given y )$, where $d$ the measurement of a physical quantity $y$. In Bayesian analysis, the posterior, which fully describes the outcome of the analysis, is the product of the likelihood and the prior. The information brought to the analysis by the data comes from the likelihood. Least-squares (LS) analysis is a consequence of using the normal (Gaussian) distribution for the likelihood. The more general likelihood analysis, from which LS analysis is derived, is not so restricted; other forms for the likelihood are permissible. 

Nuclear physics experiments offer substantial evidence of disagreement among repeated measurements. One indication is that the minimum $\chi^2$ in data analyses is often significantly larger than the number of data points. In that situation, the analyst typically inflates the final uncertainty to cover the range of dispersion of the data. The large value of $\chi^2$ indicates that either the data do not match the fitted model and/or the stated standard errors are too small and/or the assumption of a normal distribution is incorrect.

The fundamental question is, how well do experimentalists estimate the uncertainties in their results? The approach adopted here to answer that question is to examine a collection of repeated measurements and see what can be inferred about the distribution of measurements relative to their estimated uncertainties. 

Lifetime measurements of elementary particles are useful for this purpose because there are approximately twenty repeat measurements for each of the longest known particles. I will show that these data strongly support using a Student t distribution for the shape of the likelihood function instead of a normal distribution. Furthermore, experimenters tend to underestimate the  uncertainties in their results. It is shown that analyses based on long-tailed likelihoods, like the t distribution, gracefully cope with outlying data.

\section{Particle lifetime data}

In 1957 Gell-Mann and Rosenfeld \cite{Gellmann57} published an authoritative review of the properties of elementary particles. Their work quickly led to the formation of the Particle Data Group, which now summarizes the known properties of elementary particles on an annual basis. For each particle property, the committee: 
(a) lists all relevant experimental data,
(b) decides which data to include in its final analysis (outliers often rejected),
and (c) states the best current value and its estimated standard error. 
The final results are typically obtained using the least-squares average of the accepted data. The standard error is often magnified by $\sqrt{\chi^2/(n-1)}$ to take into account the dispersion of the data. 

Incidentally, in half of all 64 PDG tables involving three or more entries the standard error is adjusted, and when it is, the average scale factor is 2.0. These numbers indicate the frequent occurrence of significant disagreements among particle-physics measurements, relative to their quoted uncertainties. These observations are even more remarkable because the data involved have been carefully selected by the PDG. 

The PDG reports \cite{PDG06} are an excellent source of information about measurements of unambiguous physical quantities. They are available online, and provide insight into how physicists interpret data.

\begin{figure} 
\begin{tabular}{c} 
  \includegraphics[width=7cm]{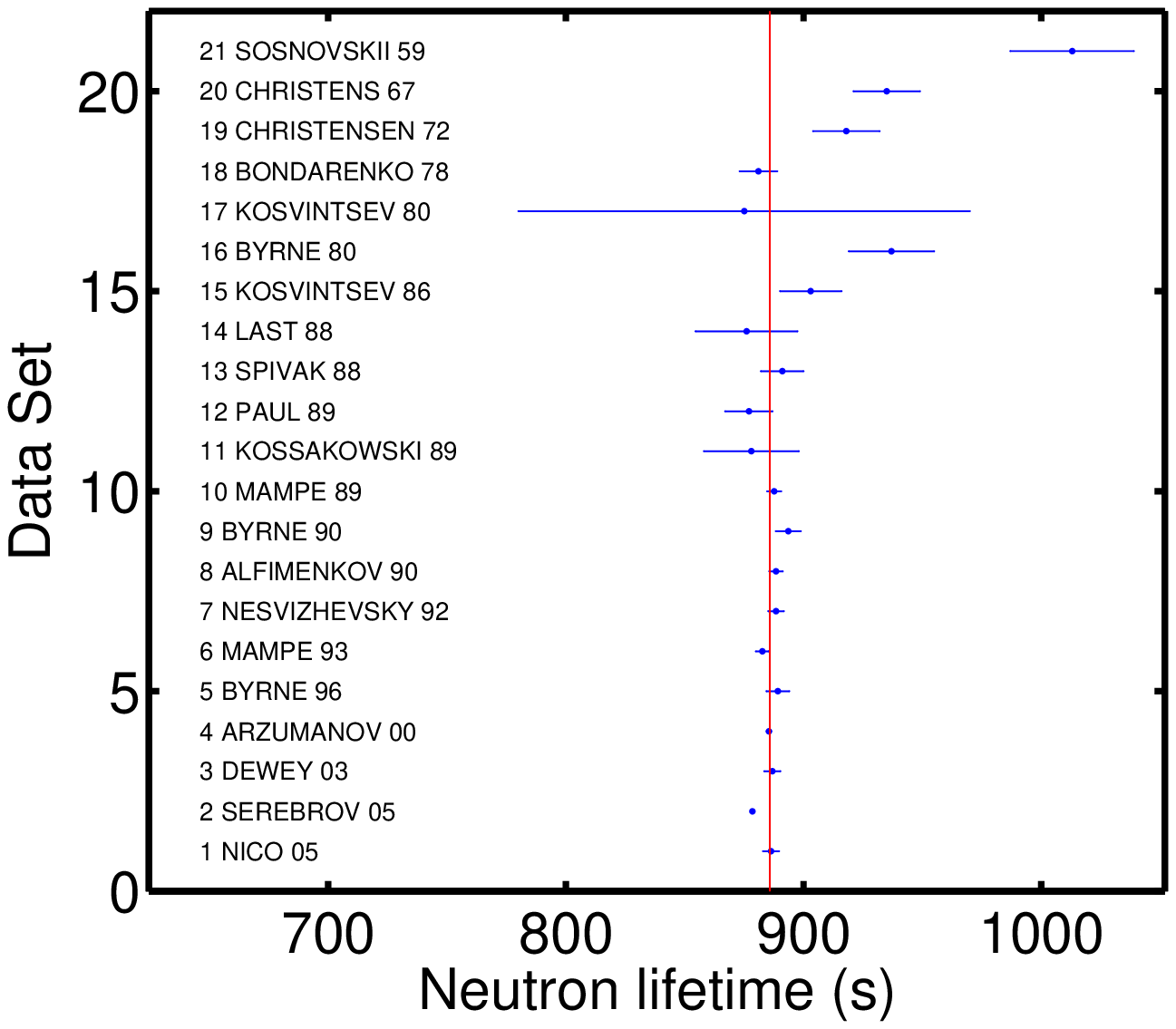} 
  \includegraphics[width=7cm]{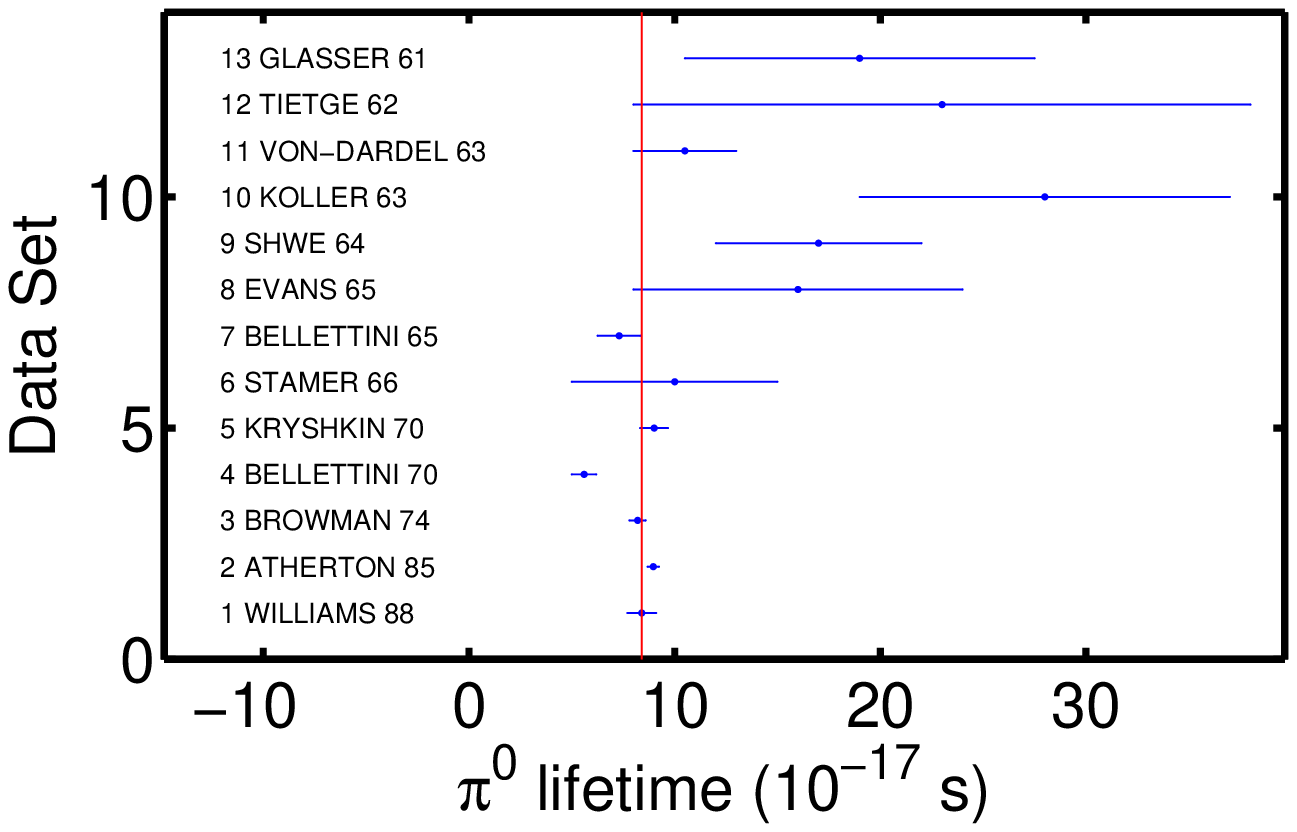}
\end{tabular}  
  \caption{ Plots of all measurements, in chronological order, of the lifetimes of the neutron and $\pi^0$ particles.   }
\label{fig:ndata}
\end{figure}

Figure~\ref{fig:ndata}a shows all measurements of the lifetimes of the neutron compiled from Ref.~\cite{PDG06} and earlier PDG reports.  
The vertical line is the PDG value, which includes the seven most recent data sets, except for \#2 \cite{Serebrov05}, because it disagrees with the rest by 9.5 standard errors. For all 21 points, $\chi^2$ (relative to PDG) = 149. It is clear from the plot that the data set contains several outliers, that is, measurements that disagree with the PDG value by more than three times their stated standard errors. 

Figure~\ref{fig:ndata}b shows all measurements of the lifetimes of the $\pi^0$ meson. The PDF value, indicated by the vertical line, is based on the four most recent values, excluding the latest one (\#1). Included in the PDG average is \#4, which is 4.7 standard errors away from the PDG average. For the 13 data points, $\chi^2$ (relative to PDG) = 40. 

\begin{figure} 
\begin{tabular}{cc}
  \includegraphics[width=7cm]{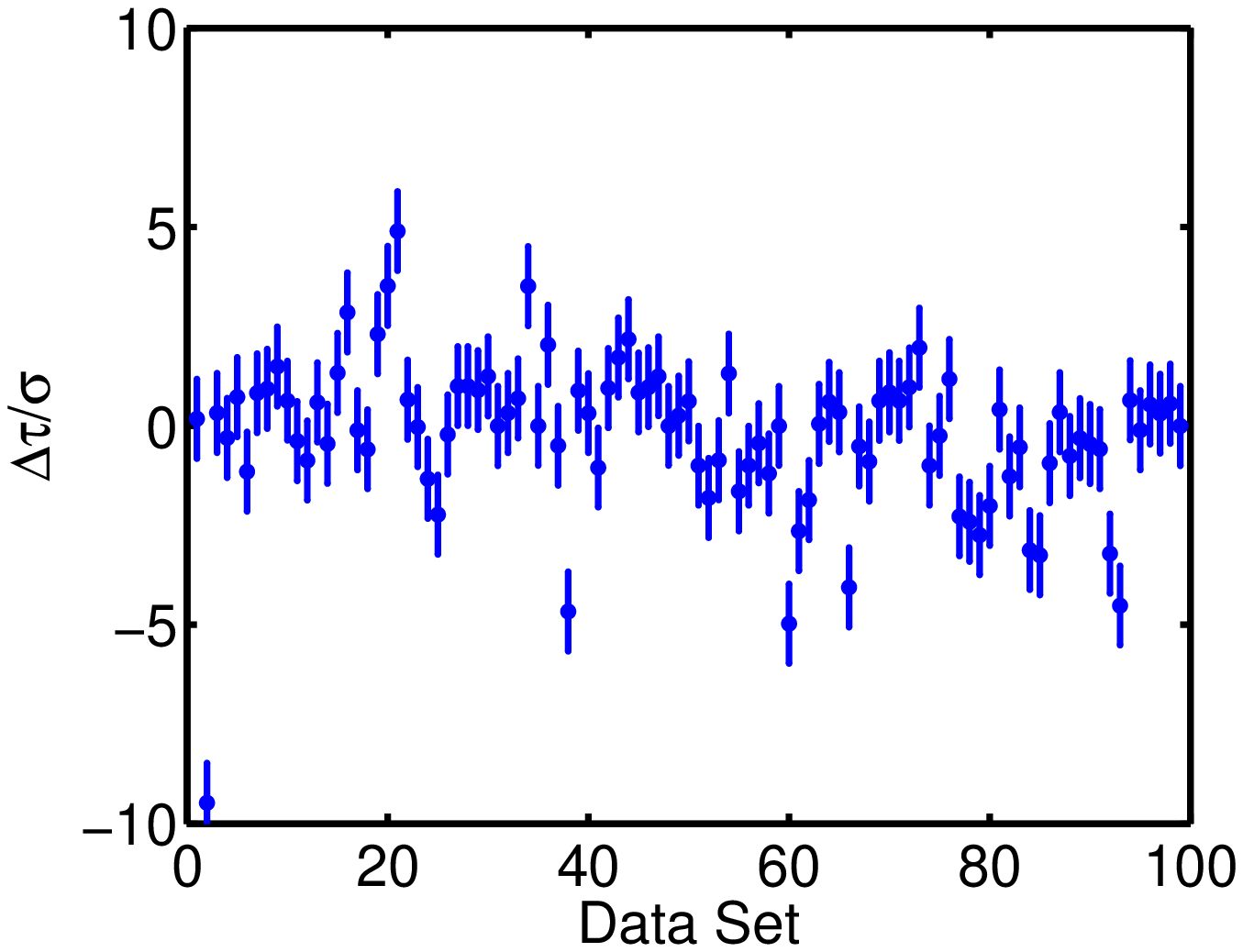} 
  \includegraphics[width=7cm]{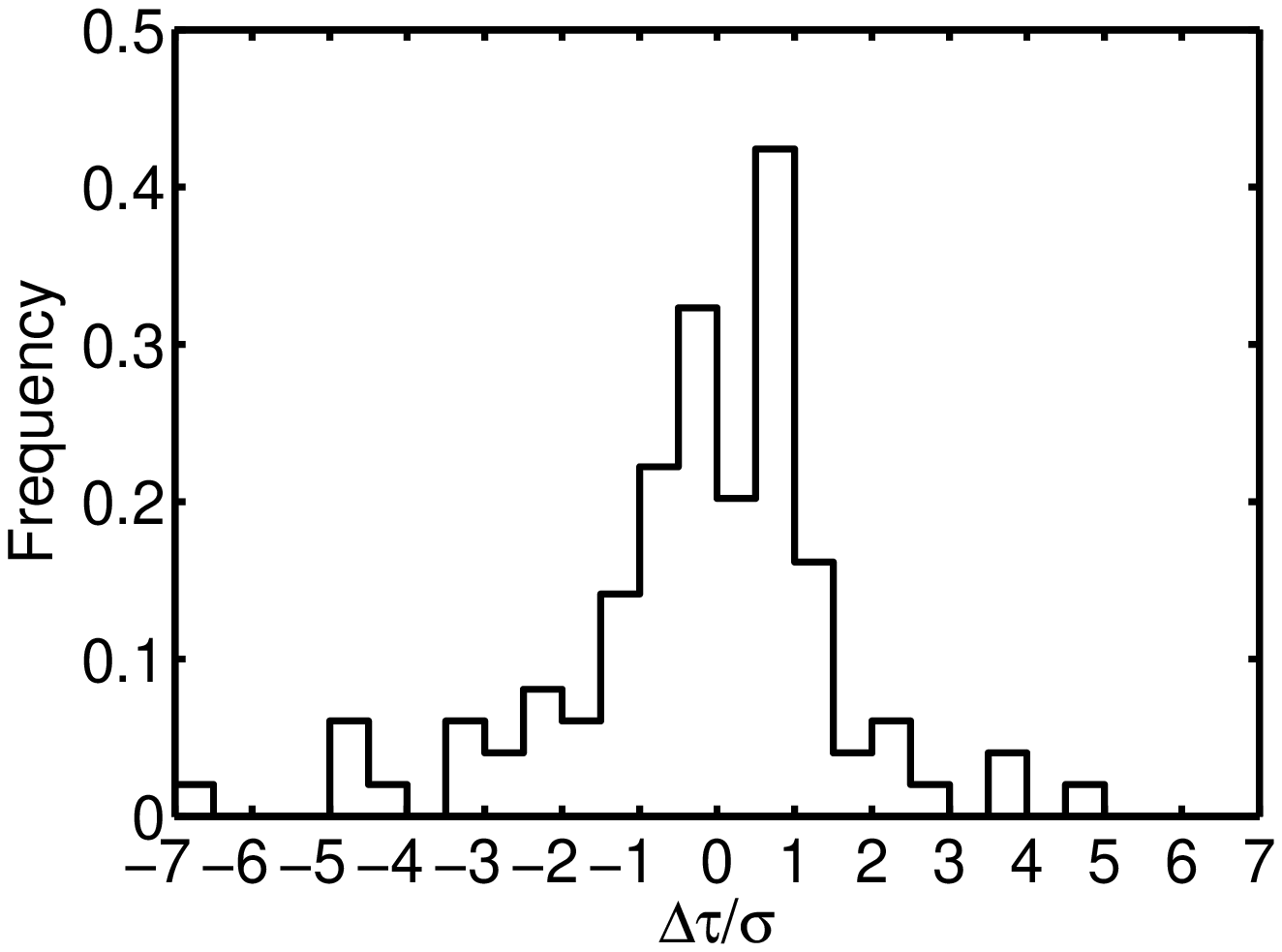}
\end{tabular}  
  \caption{(a, left) Composite of lifetime measurements of five elementary particles, n, $\mu^\pm$, $\pi^0$, $K^0_s$, and $\Lambda$. The discrepancy of each measurement from the recent PDG value is divided by the quoted standard error. (b, right) Histogram of the normalized discrepancies.}
\label{fig:ltresids}
\end{figure}

Figure~\ref{fig:ltresids}a shows the discrepancies of 99 lifetime measurements for five particles from PDG values, divided by their standard errors, $\Delta\tau/\sigma$. These five particles are chosen because they were among the first discovered of the unstable particles. 
Figure~\ref{fig:ltresids}b shows the histogram of these normalized discrepancies. The objective of the present study is to characterize the distribution of discrepancies relative to their estimated uncertainties, $\Delta\tau/\sigma$. For these 99 data points, $\chi^2$ = 367, indicating their rms fluctuation is twice their stated standard errors.

For a first cut at analyzing any data set, the analysis suggested by John Tukey\cite{Tukey77} is useful. The steps are: 
(a)~find the quartile positions in the data set, Q1, Q2, Q3;
(b)~calculate the inter-quartile range, IQR = Q3 - Q1; 
(c)~determine the fraction of data in the intervals $y$ < Q1 - 1.5 IQR and $y$ > Q3 + 1.5 IQR, called the suspected outlier fraction (SOF). For the normal distribution, IQR = 1.35 $\sigma$ and SOF = 0.7\%.
The IQR is a measure of the width of the core of the distribution and the SOF the extent of its tail, relative to the width of the core. Q2 is the median, of course, which is a good estimate of the measurand.

For the composite lifetime data shown in Fig.~\ref{fig:ltresids}a, IQR is 1.83. Thus, the width of the core of this distribution is 1.36 times larger than the value 1.35 that would obtain if the distribution were normal and the standard errors were correctly estimated. Furthermore, SOF = 6.6\%, indicating this distribution has about ten times as many data in its tails than expected for a normal, denoting a distribution with long (fat) tails. 

\section{Uncertainties in physics experiments}

When an experimenter states his/her measurement of a physical quantity $y$ as $y = d \pm \sigma$, the standard error $\sigma$ represents experimenter's estimated uncertainty in $d$. This statement is interpreted probabilistically as a likelihood function, the conditional probability of $d$, given the measurand value $y$ and the stated standard error $\sigma$, $p(d \given  y \,\sigma \, I)$,
where $I$ represents any relevant background information, for example, how the experiment is performed. The likelihood is a probability density function (PDF) in the variable $d$, so it is properly normalized to unit area with respect to $d$.
However, the likelihood is usually viewed as function of $y$, and is not necessarily normalized wrt. to $y$.  The likelihood is usually taken to be a normal distribution (Gaussian) with standard deviation $\sigma$.

Experimental uncertainties are usually thought of as consisting of two types. The first type is statistical uncertainty, which often arises from noise in the measured signals or events being counted. In the latter case, the uncertainty is usually estimated using the Poisson distribution. In the former, the rms fluctuations in the signals can be measured. These sources of uncertainty are usually Type A uncertainties, that is, they can be quantified by repeated measurements and estimated using frequentist statistical methods. Statistical uncertainties are likely to be estimated reliably. 

The second type is called systematic uncertainty because it often affects many or all of the experimental results from an experiment.
These arise from uncertainties in equipment calibration, experimental procedure, or corrections to the data. In nuclear physics, typical systematic uncertainties arise from detector efficiencies and deadtimes, target densities and thickness, and integrated beam fluence. Systematic uncertainties are
usually Type B uncertainties; they are often determined by nonfrequentist methods, and may be based on the experimenter's judgment. Hence, these uncertainties may be subjective, difficult to assess, and possibly not well known.      

Statistical and systematic uncertainties are usually added in quadrature (rms sum).


\subsection{Uncertainty in the uncertainty}
\label{sec:uncert}

Suppose there is uncertainty in the stated standard error $\sigma_0$ for measurement $d$. 
Dose and von der Linden \cite{Dose99} presented the following plausible derivation of a suitable likelihood function.
They assume the likelihood has an underlying normal distribution,
\begin{equation}  
\label{eq:normal}     
p(d \given y \,\sigma \, I) \propto \exp\left[ - \smallfrac{1}{2} \left( \frac{d - y}{\sigma}\right)^2 \right] \; , 
\end{equation}
where $\sigma$ is the standard deviation of the distribution. Because the experimenter's stated standard error $\sigma_0$ is uncertain, $\sigma$ is assigned a probability density function (PDF). Rather than working directly with $\sigma$, Dose and van der Linden consider the variable $\omega$, where $\sigma$ is scaled by $\sigma = \sigma_0/\sqrt{\omega}$, for which they assign a Gamma distribution:
\begin{equation}  
\label{eq:pomega}     
p(\omega \given \, I) \propto \omega^{a - 1}\exp\left( - a \, \omega \right) \; . 
\end{equation}
The mean of this distribution is $\omega = 1$ and its variance is $1/a$. The value of $a$ should be based on how uncertain one is in the uncertainty quoted by the experimenter. As $a$ approaches infinity, $p(\omega \given I)$ approaches a delta function at $\omega = 1$. When viewed as a PDF in $s$, $p(s \given I)$ has a plausible shape, with 
means of 1.59, 1.32, and 1.09, and 
rms deviations of 1.06, 0.69, and 0.30 at $\nu$ =1, 3, and 9, respectively. 

The advantage of the above parameterization is that the integration over $\omega$ can be done analytically, resulting in the likelihood   
\begin{equation}  
\label{eq:tlike}     
p(d \given y \, \sigma_0 \, I) \propto \left[ 1 + \frac{1}{\nu}\left(\frac{d - y}{\sigma_0}\right)^2 \right]^{-\frac{\nu + 1}{2}} 
\propto t_{\nu}\left( \frac{d - y}{\sigma_0}\right)\; , 
\end{equation}
which is a Student t distribution of order $\nu = 2a$.

See Refs.~\cite{Box68,Dawid73,OHagan79,Froehner89,Hanson96a,Sivia96b,Press97,Hanson05a} for other contributions to the outlier discussion.

\subsection{Properties of Student t distributions}
\label{sec:tdistr}

\begin{figure} 
\begin{tabular}{cc}
  \includegraphics[width=7cm,trim=0 0 20 20,clip]{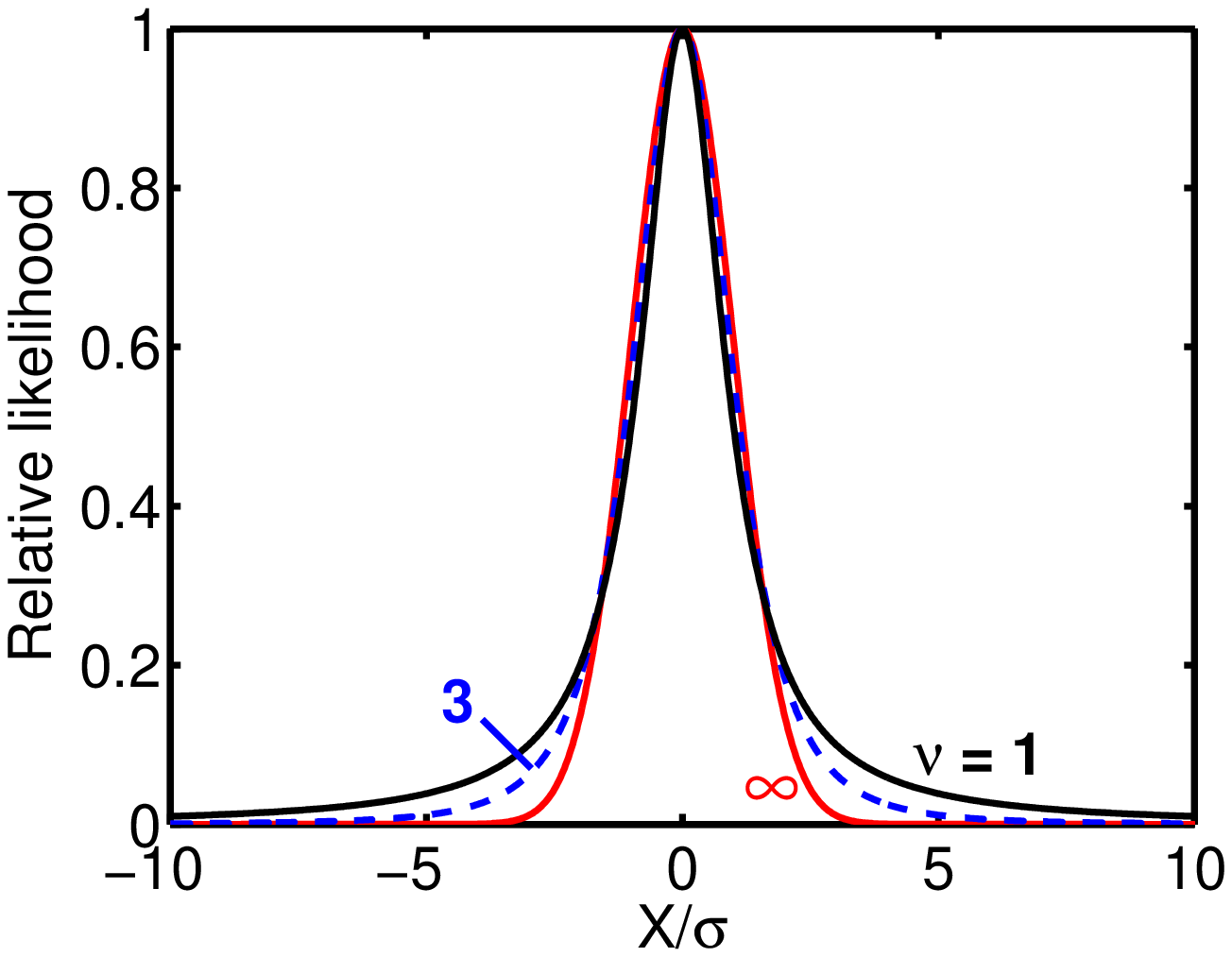} \hspace{1em} 
  \includegraphics[width=7cm,trim=0 0 10 20,clip]{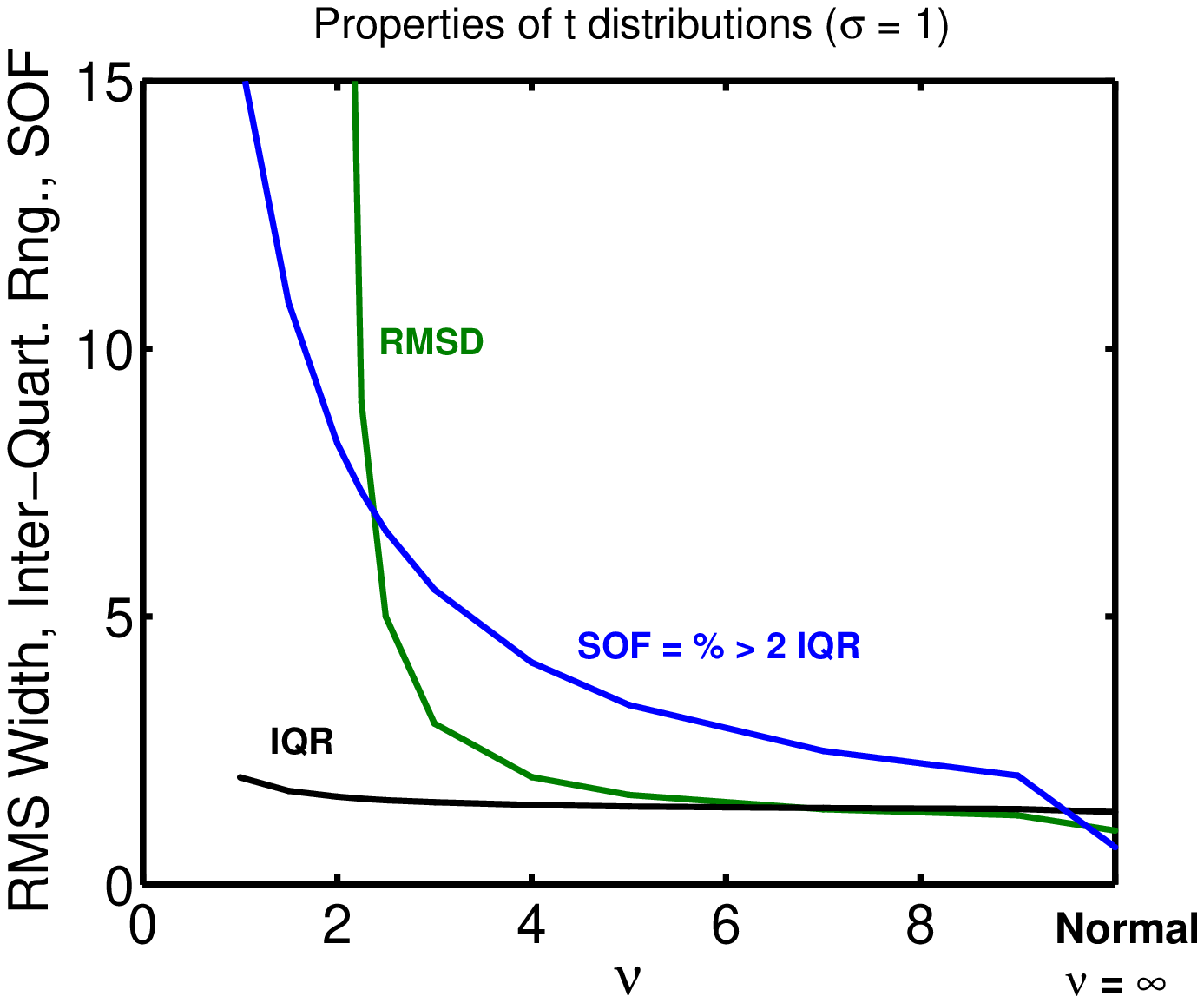} 
\end{tabular}  
  \caption{(a, left) Plots of t distributions for orders $\nu$ = 1, 3, and $\infty$ (normal). (b, right) Properties of t distributions as a function of $\nu$, the rms deviation, the IQR, and the SOF. For $\nu < 9$, SOF > 2\%. The rms deviation diverges at $\nu$ of 2 and below.}
\label{fig:tdistr}
\end{figure}

Figure~\ref{fig:tdistr}a shows the Student\footnote{Student (1908) was pseudonym for W.S. Gossett, who was not allowed to publish under his own name by his employer, Guinness brewery } $t_\nu$ distribution for three $\nu$ values, $\nu$ = 1, 3, and $\infty$. For $\nu$ = $\infty$, the t distribution is the same as the normal distribution. For $\nu$ = 3, it is the Cauchy distribution, also called the Breit-Wigner or Lorentzian in physics. The tails of $t_\nu (x)$ asymptotically fall as $|x|^{- (\nu + 1)}$. Because of that, the mean of the t distribution does not exist for $\nu \leq$ 1, and its variance is infinite for $\nu \leq$ 2. 

Figure~\ref{fig:tdistr}b displays the strong dependence of the rms width of $t_\nu (x)$ as a function of $\nu$. The Tukey quantity, Inter-Quartile Range (IQR), does not change much with $\nu$. However, the Suspected Outlier Fraction (SOF) is already 2\% at $\nu$ = 9, and increases sharply as $\nu$ drops below 4. A good first guess as to what $\nu$ matches a given data set can be  obtained from this curve and the SOF for the data set. 

\subsection{Coping with outliers}
\label{sec:outliers}

\begin{figure} 
\begin{tabular}{cc}
  \includegraphics[width=7cm]{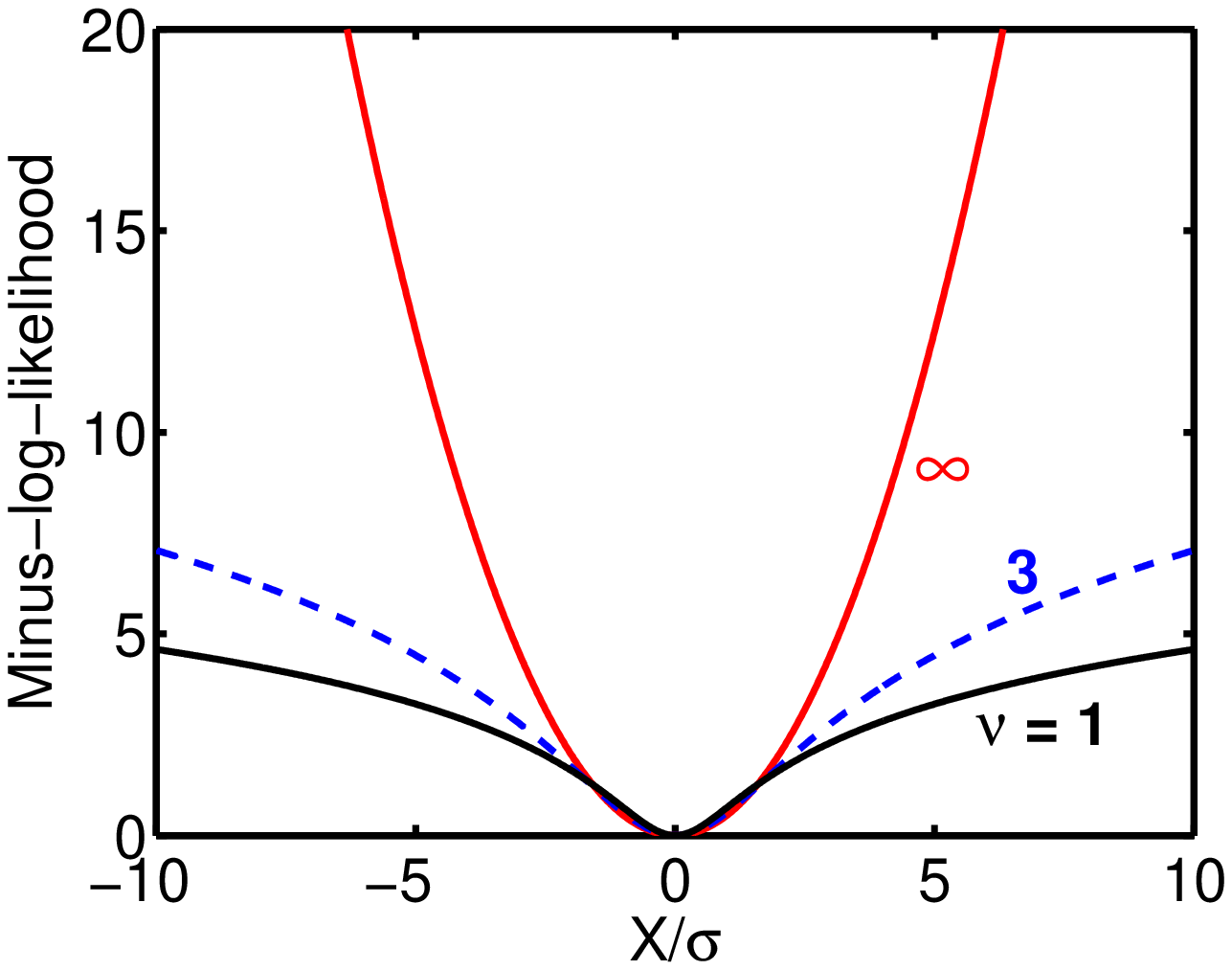} 
  \includegraphics[width=7cm]{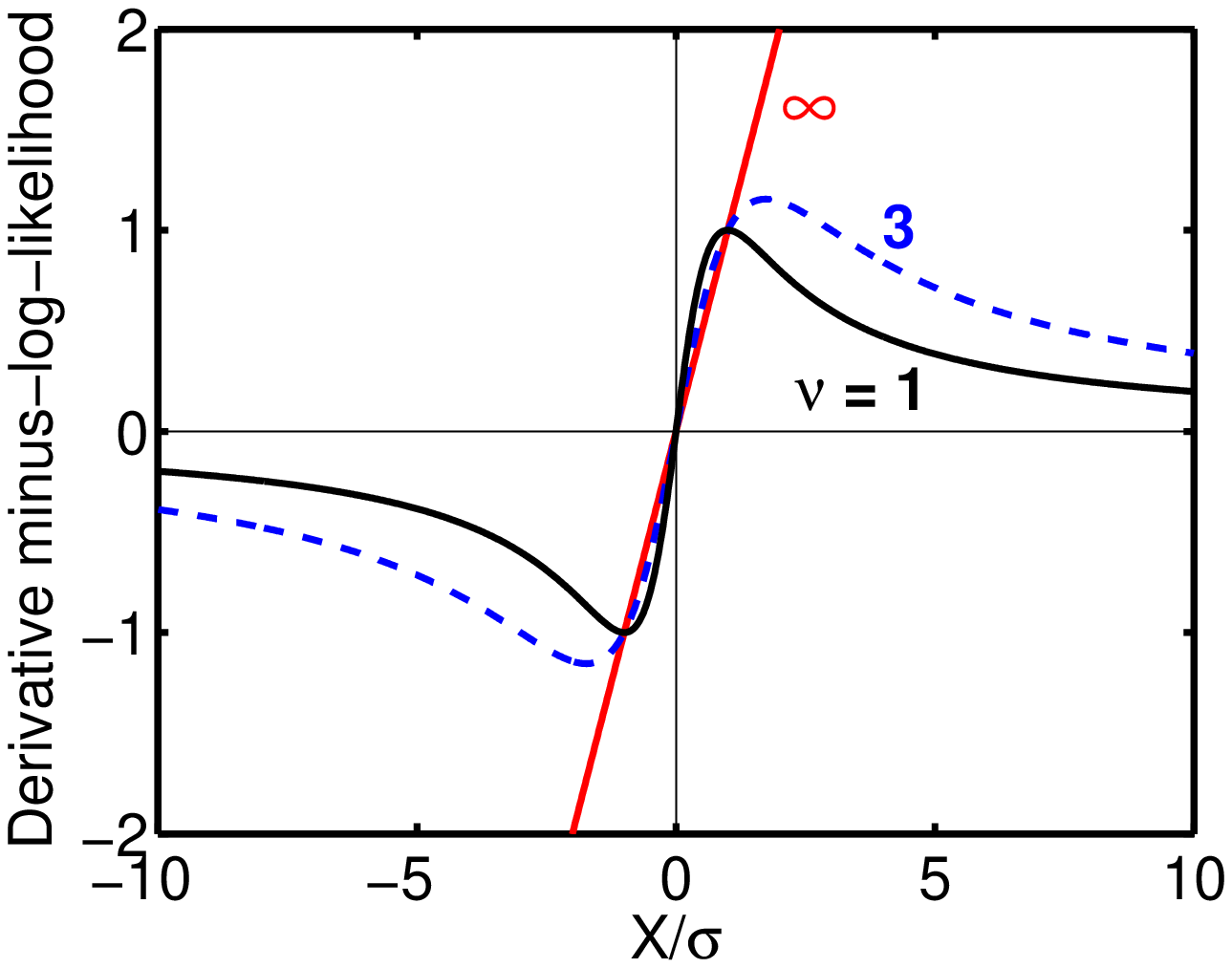} 
\end{tabular}  
  \caption{(a, left) Minus-log-likelihood dependence on the normalized residual for t distributions of order $\nu$ = 1, 3, and  $\infty$, the latter being identical to the normal distribution.  (b, right) Derivatives of the same functions. For small $\nu$ values, the restoring force of t-distribution likelihood functions drops off at large residuals, as opposed to the linearly increasing force of the normal distribution. }
\label{fig:tdistrderiv}
\end{figure}

The general Bayesian approach for coping with outliers is to use for the likelihood function a long-tailed, sometimes called a fat-tailed, distribution. Experience shows us that the exact form of the tail is not very important for ameliorating  the effect of outliers. For examples of long-tailed likelihoods and their response to outliers, see \cite{Hanson96a,Sivia96b,Press97,Dose99,Hanson05c}. 

The easiest way to understand how long-tailed likelihood functions deal with outliers is to employ the useful analogy between $\varphi(\Delta x)$ = minus-log-likelihood and a physical potential. Since the gradient of a potential is a force, $-\nabla \varphi$ is interpreted as the force with which the datum pulls on the model.  
Figure~\ref{fig:tdistrderiv}b shows the behavior of the derivative of the minus-log-t distribution. The slopes of the three curves at $x = 0$ are different, but that is not an issue here because of the scaling factor $s$ used in the present analysis.

For likelihoods to be tolerant of outliers, the restoring force eventually decreases, or at least saturates, for increasing residuals. The extent to which a likelihood function accommodates outliers can be deduced from how fast the derivative of its logarithm falls off for large residuals. The normal distribution is not outlier tolerant because it pulls ever more strongly on the solution as its residual increases. 

\section{Results}

\subsection{Analysis of composite data set}
This section summarizes the results of analyzing of the full data set composed of the lifetime measurements for the five particles, shown in Fig.~\ref{fig:ltresids}. The goal is to determine whether a t distribution appropriately describes the distribution of measurements.

From the derivation above, the likelihood for each datum is given by the t distribution \ref{sec:outliers}
\begin{equation}  
\label{eq:nlike}     
p(d_i \given \tau \, s \, \sigma_i \, I) \propto \left[ 1 + \frac{1}{\nu}\left(\frac{d_i - \tau}{s \, \sigma_i}\right)^2 \right]^{-\frac{\nu + 1}{2}}  , 
\end{equation}
where the variable $s$ has been introduced to scale all the uncertainties in the data set.
The likelihood for a data set with $n$ data points is the product of $n$ such factors, $p(\mvec{d} \given \tau \, s \, \pmb{\sigma}) \propto \prod_{i=1}^{n} \, p(d_i \given \tau \, s \, \sigma_i)$, assuming the measurements are statistically independent\footnote{An alternate approach is to infer the properties of the likelihood function is to directly fit the histogram, as in Fig.~\ref{fig:ltresids}b, using a multinomial distribution for the likelihood of each bin count. However, in the limit of infinitely narrow histogram bins, it is easy to show that this approach is equivalent to the one used here.}.
The posterior for the lifetime is obtained by marginalizing the joint distribution for $\tau$ and $s$ over the nuisance parameter $s$ 
\begin{equation}  
\label{eq:smarg}     
p(\tau \given \mvec{d} \, \pmb{\sigma}) =  
\int p(\tau \, s \given \mvec{d} \, \pmb{\sigma}) \; \mbox{d}s = 
\int p(\mvec{d}  \given \tau \, s \, \pmb{\sigma}) \, p(\tau \, s) \; \mbox{d}s\; . 
\end{equation}
The dispersion of the data points is incorporated in this posterior, much in the same way as with the $\chi^2$ scaling often applied in LS analysis. 

\begin{figure} 
\begin{tabular}{cc}
  \includegraphics[width=8.3cm,trim=0 0 10 20,clip]{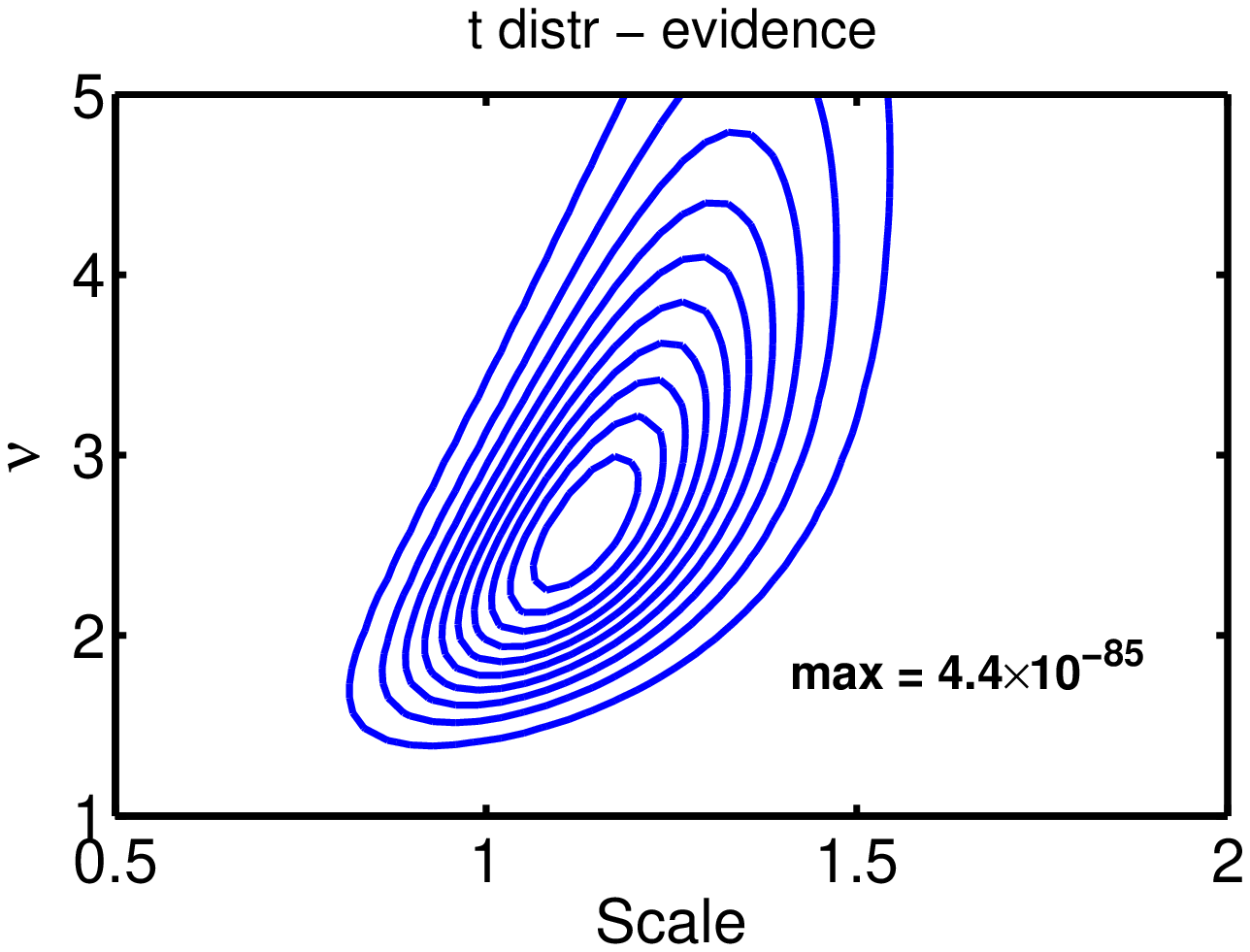} 
  \includegraphics[width=7cm]{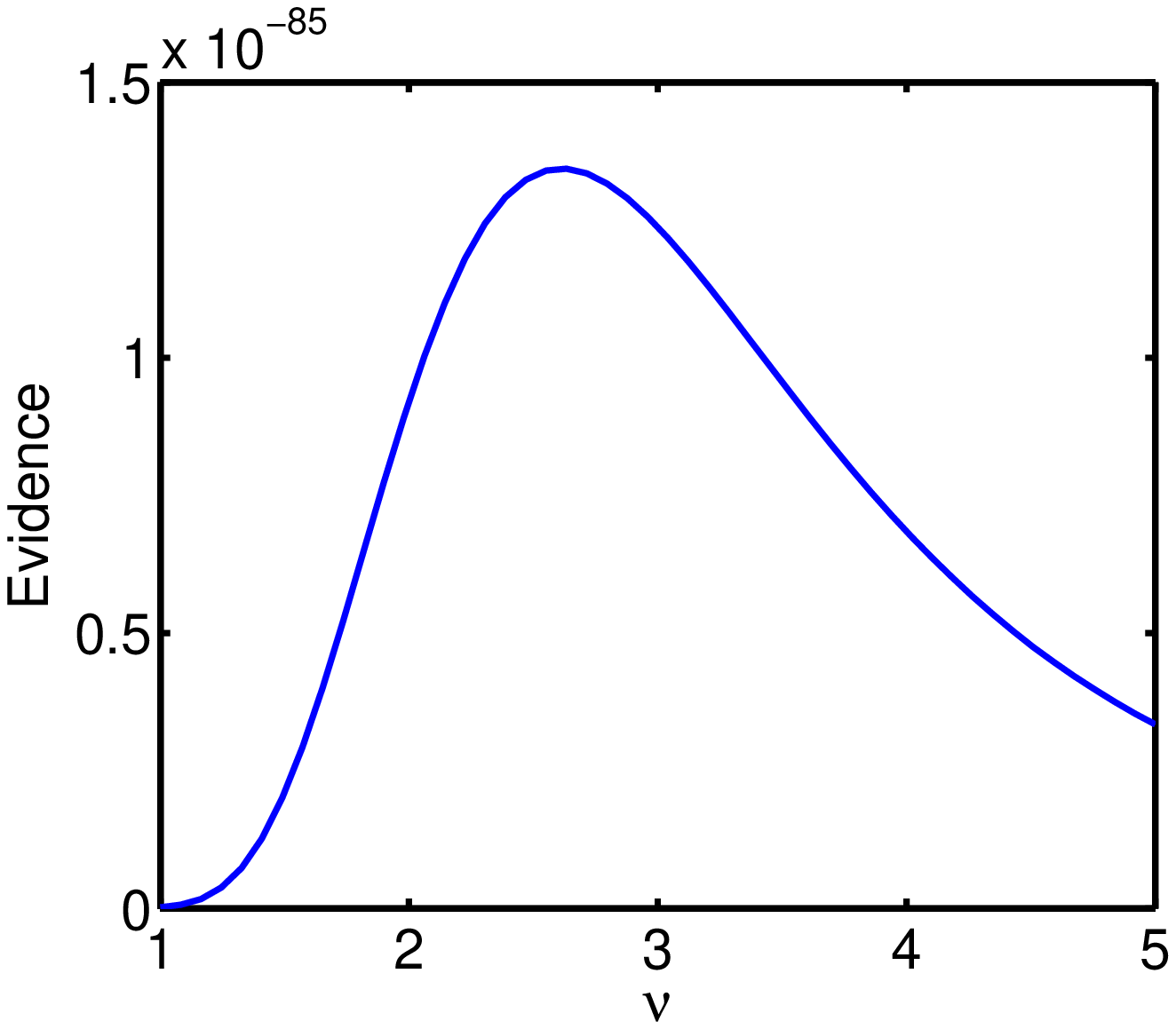} 
\end{tabular}  
  \caption{(a, left) Contour plot of the evidence as a function of order $\nu$ and the scale, $s$. (b, right) The distribution in $\nu$ has a broad peak with its maximum at $\nu = 2.6$.}
\label{fig:evidence1}
\end{figure}

To select between the two models, the t or normal distribution (T or N), Bayes rule \cite{Sivia06} gives the odds ratio as  
\begin{equation}  
\label{eq:modelodds}    
\frac{p(\mbox{T} \given \mvec{d}\,\pmb{\sigma} \, I)}
{p(\mbox{N} \given \mvec{d}\,\pmb{\sigma} \, I)} = 
\frac{p(\mvec{d} \given  \mbox{T}\, \pmb{\sigma} \, I)}
{p(\mvec{d} \given  \mbox{N}\, \pmb{\sigma} \, I)} 
\frac{p(\mbox{T} \given \, I)}
{p(\mbox{N} \given \, I)}\; , 
\end{equation}
where ${p(\mbox{T} \given I)}/{p(\mbox{N} \given I)}$ is ratio of the priors for the two models,
and $p(\mvec{d} \given  \mbox{T}\, \pmb{\sigma} \, I)$ is the evidence. The evidence is evaluated as the integral over the joint distribution in $\tau$ and $s$
\begin{equation}  
\label{eq:evidence}     
p(\mvec{d} \given  \mbox{T}\, \pmb{\sigma} \, I) = \int\ p(\mvec{d} \given  \tau \, s \, \mbox{T}\, \pmb{\sigma} \, I) \; p(\tau \, s  \given  \mbox{T}\, \pmb{\sigma} \, I) \; \mbox{d}\tau  \, \mbox{d}s \; , 
\end{equation}
shown in Fig.~\ref{fig:evidence1}a.
In this study, the evidence integral is estimated from the integrand values evaluated on a uniform grid. The maximum in the evidence in Fig.~\ref{fig:evidence1}b occurs at $\nu\approx 2.6 \pm 0.9$. The odds ratios of the t distribution (at $\nu = 2.6$) to the normal is $1.3\times10^{-85} /2.2\times10^{-90} = 5.5\times10^4$,
assuming the odds ratio of the priors on the models is unity and the priors on the parameters are the same for both models.
The priors on $\tau$, $s$, and $\nu$ are constant. 
Thus, the t distribution is strongly preferred by the data to the normal distribution. 
For the normal distribution, $\bar{s} = 1.95 \pm 0.14$.
; $s = 1$ is definitely rejected. 

Excluding the data point with the largest discrepancy (9.5 sigma) yields an odds ratio of 
$2.1\times10^{-82} /5.0\times10^{-84} = 42$, which is considerably less than above, but the t distribution is still preferred over the normal. In this case, the scaling for the normal distribution is $\bar{s} = 1.71 \pm 0.12$.

\subsection{Analysis of the neutron lifetime measurements}

\begin{ltxfigure}
\begin{minipage}{0.48\textwidth}
\centering
  \includegraphics[height=5.5cm,trim=0 0 10 20,clip]{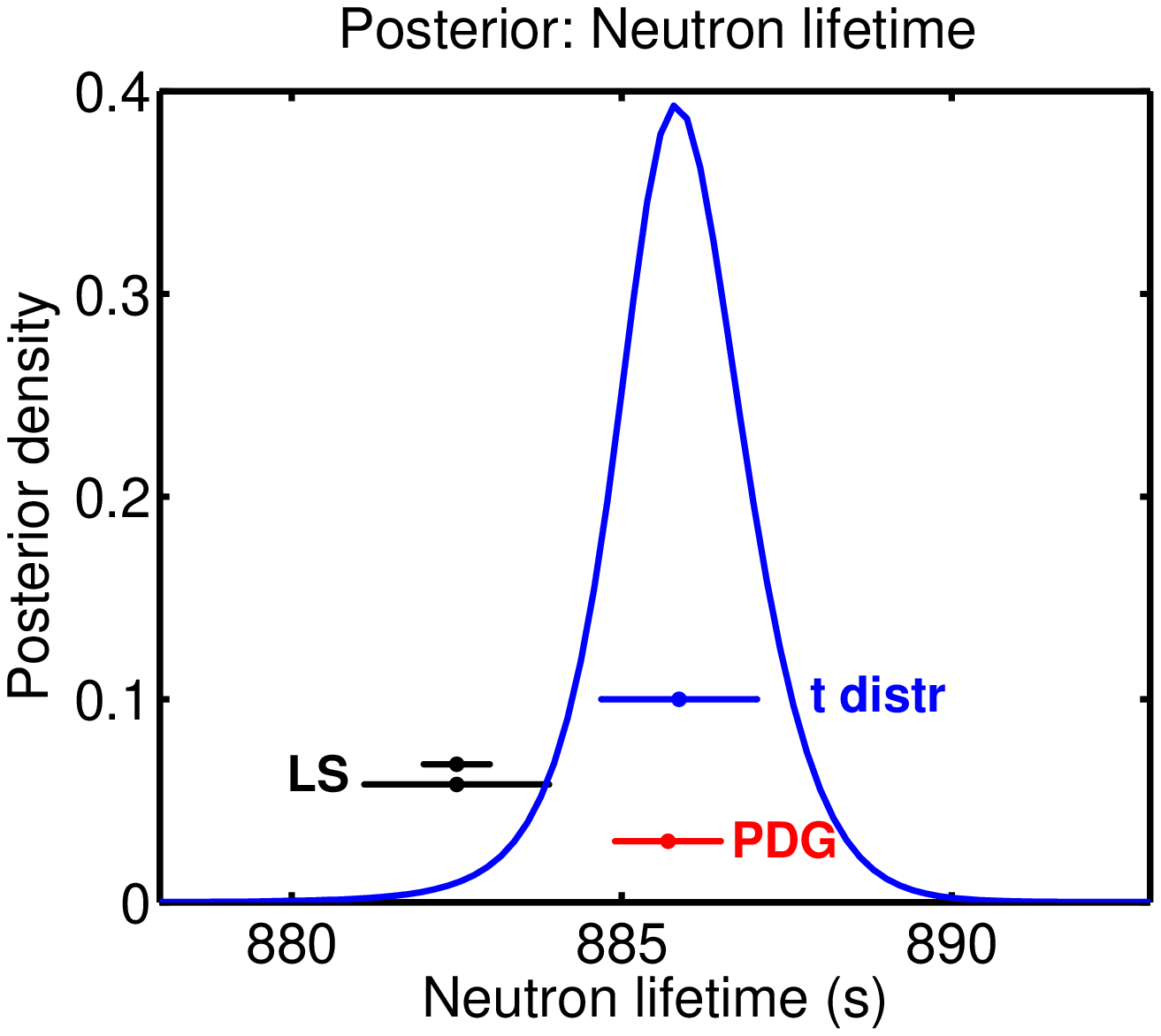}
  \caption{Posterior distribution for the t-distribution analysis of the neutron-lifetime data. The estimated lifetime and its standard error are shown for the t analysis and LS, based on the full set of 21 data. The PDG estimate is based on combining the seven most recent measurements, excluding that of Serebrov et al. }
  \label{fig:nposterior} 
\end{minipage}%
\hfill
\begin{minipage}{0.48\textwidth}
\centering    
  \includegraphics[height=5.5cm,trim=0 0 10 20,clip]{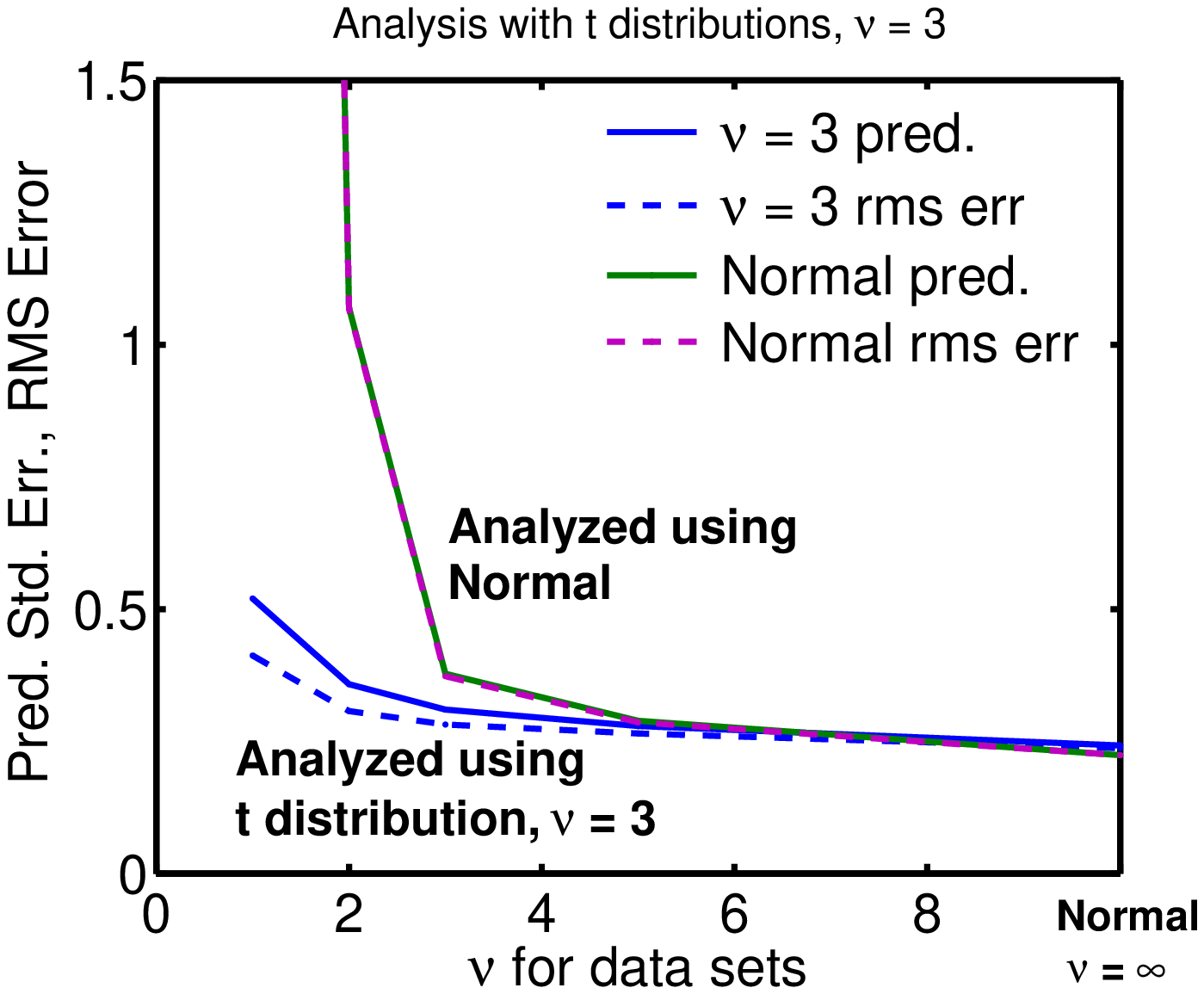} 
  \caption{Results of tests comparing the performance of analyses based on t-distribution likelihoods to those based on using the normal distribution. The t-distribution analysis does much better than the normal for data sets from small $\nu$, although the predicted or estimated uncertainties are slightly larger than the actual rms error.}
  \label{fig:ttest}
\end{minipage}%
\end{ltxfigure}

Figure~\ref{fig:nposterior} shows the result of analyzing all 21 measurements of the neutron lifetime shown in Fig.~\ref{fig:ndata}a using for the likelihood t-distributions with $\nu = 2.6$. The posterior in $s$ is obtained by marginalizing the joint distribution in $\tau$ and $s$ over $s$, as mentioned above. The posterior distribution for $s$ has a mean value of $\bar{s} = 1.16$, indicating a fairly small scaling factor is required when using a t-distribution likelihood. By contrast, the least-squares result, shown in Fig.~\ref{fig:nposterior} with $\chi^2$ scaling, applies a scaling factor of $s = 2.73$. This large scaling of the uncertainty is required because the data are more disperse than a normal distribution with the quoted standard errors.

The PDG estimate, based on the seven most recent measurements, excluding the one by Serebrov et al. \cite{Serebrov05} is reasonably consistent with the present result, although its standard error is somewhat smaller because of the relative good consistency of the seven chosen data. The standard error from the t-distribution analysis lies between that of the LS and PDG results, while deftly coping with the Serebrov outlier.


\subsection{Tests of performance of t-distribution analysis}

How well does the analysis based on a t distribution of a given order work for data drawn from different distributions? To answer that question, a series of Monte Carlo tests are performed.  
In these tests, 10000 data sets are created, each with 20 data values drawn from a specific t distribution. Each data set is analyzed with a likelihood function, either the t distribution of order $\nu = 3$ or the normal, which is equivalent to a least-squares (LS) analysis. The results from all data sets are summarized in terms of the mean estimate and its rms error. The rms estimated standard error is also calculated.  

Figure~\ref{fig:ttest} shows the results of these tests. The performance of the LS (normal) analysis on data drawn from t distributions with $\nu \leq 3$ is poor to very bad. This result is expected because these data tend to have a number of outliers. The analysis based on the t distribution with $\nu = 3$  excels when the data contain a significant fraction of outliers, that is, when they are drawn from t distributions with $\nu \leq 3$. 
For normally distributed data ($\nu = \infty$), however, the LS analysis has slightly smaller rms error than the t-distribution analysis by about 4\%. 

To summarize these results, LS analysis exposes the analyst to dire consequences in the presence of outliers in the data. On the other hand, using likelihoods based on t distributions with $\nu \approx 3$ gracefully deals with outliers while achieving very close to the same accuracy as the LS analysis when the data are normally distributed.      

\section{Discussion}

The present study demonstrates that particle lifetime data are much better described by a likelihood function based on the Student t distribution with $\nu \approx$ 2.6 to 3.0 than by a normal distribution.
Furthermore, likelihood or Bayesian analyses based on the t distribution cope well with outliers, while treating each datum in the same way. There is no need to identify outliers and specially deal with them. Furthermore, t-distribution analysis produces stable results when outliers exist in data sets, whereas the normal distribution does not. These results for particle lifetimes do not represent all physical measurements, but are worth keeping in mind.

A useful conclusion of this study is that repeat experiments are worthwhile to gain confidence and mitigate against outliers, even though they might not substantially improve on the accuracy of earlier experiments. While the use of t distributions reduces the influence of outliers, when an outlier is detected, every effort should be made to try to understand the details of the experiments and possible causes for the disagreements.

As a word of caution, long-tailed likelihood functions may result in posteriors with multiple maxima, which may complicate the analysis. 
While the posterior mean is the best estimator, it can be computationally costly to evaluate, especially for nonlinear models with many parameters.

It is expected that the experimental uncertainty in a given data set may contain statistical components that follow normal (or Poisson) distribution and systematic uncertainties that potentially follow t distributions. In that case, the likelihood is a convolution of normal and t distributions, which can not easily be represented analytically.

Some outlier models \cite{Sivia96b,Dose99} adopt the notion that the data set contains both good data and bad data. The likelihood is written as a mixture of normal and t distributions (or other long-tailed function), for example, $(1 - \beta) \, \mbox{N} + \beta \, \mbox{T}$, where N stands for the normal and T the t distribution. 
This treatment allows for either T (with probability $\beta$) or N (with probability $(1 - \beta)$), which form may satisfactorily approximate the convolution of normal and t  distributions suggested in the previous paragraph. 

\section*{Acknowledgments}
The author is grateful to Anthony Hill, Frederik Tovesson, Robert Haight, John Ullmann, Toshihiko Kawano, Patrick Talou, Gerald Hale, and Ludovic Bonneau for many useful discussions. 
This work was done under U.S. DOE Contract DE-AC52-06NA25396.


\end{document}